\documentclass[reprent,aps,prb,twocolumn,
superscriptaddress,nofootinbib]{revtex4-2}
\usepackage{amsmath,amssymb}
\usepackage{graphicx}% Include figure files
\usepackage{bm}% bold math
\usepackage{multirow}
\usepackage{braket}

\usepackage{color}

\begin{document}

\title{
Field-direction-dependent skyrmion crystals in noncentrosymmetric cubic magnets: \\
A comparison between point groups $(O,T)$ and $T_{\rm d}$
}
\author{Satoru Hayami}
\email{hayami@phys.sci.hokudai.ac.jp}
\affiliation{Graduate School of Science, Hokkaido University, Sapporo 060-0810, Japan}
\author{Ryota Yambe}
\affiliation{Department of Applied Physics, The University of Tokyo, Tokyo 113-8656, Japan }

\begin{abstract}
We investigate the instability toward a skyrmion crystal (SkX) in noncentrosymmetric cubic magnets with an emphasis on a comparison between point groups $(O,T)$ and $T_{\rm d}$. 
By constructing low-temperature magnetic phase diagrams under an external magnetic field for three directions based on numerically simulated annealing, we find that the system under the point group $(O,T)$ exhibits different two types of SkXs depending on the field direction, while that under $T_{\rm d}$ does not show such an instability. 
The difference between them is understood from the difference in the momentum-dependent Dzyaloshinskii-Moriya interaction under each point group.  
Meanwhile, we show that the system under $T_{\rm d}$ leads to the SkX instability by considering an additional effect of the uniaxial strain, which lowers the symmetry to $D_{\rm 2d}$. 
We obtain two different SkXs: N\'eel-type and anti-type SkXs, the former of which is stabilized in the presence of the interactions at the three-dimensional ordering wave vectors. 
The present results provide rich topological spin textures in the three-dimensional systems, which are sensitive to the magnetic-field direction and point-group symmetry.   
\end{abstract}

\maketitle

\section{Introduction}

Topological spin textures have been extensively studied in various fields of condensed matter physics, since they give rise to intriguing low-energy excitations and physical phenomena~\cite{nagaosa2013topological, garst2017collective}. 
In solids, they often appear by superposing multiple spiral waves along different directions, which are referred to as the ``multiple-$Q$" state. 
Depending on the direction of magnetic wave vectors and the lattice structures, a variety of multiple-$Q$ states are realized~\cite{hayami2021topological}, such as the skyrmion crystal (SkX) characterized by the triple-$Q$ (double-$Q$) modulations on a triangular (square) lattice and the hedgehog crystal by the triple-$Q$ or quartet-$Q$ modulations on a cubic lattice. 
The search for the stabilization conditions of the multiple-$Q$ states has been still an active research field in both theory and experiments. 

One of the most fundamental mechanisms to stabilize the SkX is the competition between the ferromagnetic exchange interaction and the Dzyaloshinskii-Moriya (DM) interaction~\cite{dzyaloshinsky1958thermodynamic,moriya1960anisotropic} under an external magnetic field~\cite{rossler2006spontaneous, Binz_PhysRevLett.96.207202,  Binz_PhysRevB.74.214408, Yi_PhysRevB.80.054416,  Butenko_PhysRevB.82.052403}, the latter of which is characterized by the antisymmetric anisotropic exchange interaction and originates from relativistic spin--orbit interaction in the absence of the inversion center.  
The spin model incorporating these interactions provides a deep understanding of the SkX in real materials~\cite{Tokura_doi:10.1021/acs.chemrev.0c00297}, such as MnSi~\cite{Muhlbauer_2009skyrmion, Neubauer_PhysRevLett.102.186602}, Fe$_{1-x}$Co$_x$Si~\cite{yu2010real}, and Cu$_2$OSeO$_3$~\cite{seki2012observation, Adams2012, Seki_PhysRevB.85.220406} under the chiral point group $T$ and Co$_8$Zn$_9$Mn$_3$~\cite{Karube_PhysRevB.98.155120} under the chiral point group $O$. 
In addition, the spin model extended to multiple-spin interactions describes further intriguing SkXs in materials, such as the SkX with the short-period modulation in EuPtSi~\cite{kakihana2018giant,kaneko2019unique,tabata2019magnetic,kakihana2019unique,hayami2021field}. 
Meanwhile, there have been few studies for the multiple-$Q$ instability in the other noncentrosymmetric point group $T_{\rm d}$, since there are no Lifshitz invariants in the free energy~\cite{dzyaloshinskii1964theory,kataoka1981helical, Bogdanov89, Bogdanov94}; the role of the DM interaction on magnetic states under $T_{\rm d}$ has not been fully elucidated in a systematic way. 
Furthermore, an effective spin model with not only antisymmetric anisotropic exchange interactions but also symmetric ones has been recently formulated under various crystal symmetries including $T_{\rm d}$~\cite{Yambe_PhysRevB.106.174437, yambe2023anisotropic}. 
However, it has been still unclear when and how the SkX and other multiple-$Q$ states appear in the phase diagram under these systems.

In the present study, we investigate the possibility of the multiple-$Q$ states in a bilinear spin model with the DM interaction on a simple cubic lattice under the point group $T_{\rm d}$. 
To understand the role of the DM interaction originating from the $T_{\rm d}$ symmetry, we compare the result under $T_{\rm d}$ with that under chiral point groups $(O,T)$.  
The similarities and differences between $(O,T)$ and $T_{\rm d}$ are discussed by systematically constructing low-temperature magnetic phase diagrams for different magnetic-field directions in both point groups based on the simulated annealing.
As a result, we find that two types of SkXs are realized in the case of $(O,T)$, while there is no multiple-$Q$ instability in the case of $T_{\rm d}$. 
We show a way of inducing multiple-$Q$ states in the model with the $T_{\rm d}$-type DM interaction by considering the effect of the uniaxial stress so that the symmetry reduces to $D_{\rm 2d}$. 
In contrast to the Bloch-type SkX in the chiral point groups $(O,T)$, we obtain the N\'eel-type and anti-type SkXs with the opposite sign of the scalar chirality in the phase diagram. 
Our results indicate rich topological spin textures in noncentrosymmetric cubic point groups, which will be useful for the exploration of further exotic multiple-$Q$ states in three-dimensional systems.

The rest of this paper is organized as follows. 
In Sec.~\ref{sec: Model and method}, we introduce an effective spin model including the momentum-dependent DM interactions on the cubic lattice. 
We discuss the similarities and differences in the models between the point groups $(O,T)$ and $T_{\rm d}$. 
We also outline the numerical method based on the simulated annealing. 
We discuss the instability toward the SkX in Sec.~\ref{sec: Results}. 
We show the magnetic phase diagrams for three different magnetic-field directions under both $(O,T)$ and $T_{\rm d}$. 
Then, we examine the effect of the uniaxial strain under the point group $T_{\rm d}$. 
We summarize this paper and discuss the possibility of the multiple-$Q$ instability under $T_{\rm d}$ in Sec.~\ref{sec: Summary}. 

\section{Model and method}
\label{sec: Model and method}

We consider the model Hamiltonian on a simple cubic lattice with the lattice constant $a=1$, which is given by 
\begin{align}
\label{eq: Ham1}
\mathcal{H}=&-\sum_{\bm{q}}  \left[J_{\bm{q}} \bm{S}_{\bm{q}} \cdot  \bm{S}_{-\bm{q}} 
+ i \bm{D}_{\bm{q}} \cdot (\bm{S}_{\bm{q}} \times  \bm{S}_{-\bm{q}})\right] \nonumber \\
&-\sum_i \bm{H}\cdot \bm{S}_i, 
\end{align}
where the first term represents an interaction in momentum space; the summation regarding $\bm{q}$ is taken over the first Brillouin zone. 
$\bm{S}_{\bm{q}}=(S^x_{\bm{q}}, S^y_{\bm{q}}, S^z_{\bm{q}})$ is the Fourier transform of the localized spins $\bm{S}_i=(S_i^x, S_i^y, S_i^z)$ with the fixed length $|\bm{S}_i|=1$, and $J_{\bm{q}}$ and $\bm{D}_{\bm{q}}$ are the Fourier transforms of the interactions in real space, i.e., $J_{ij}$ and $\bm{D}_{ij}$, respectively. 
We suppose the absence of spatial inversion symmetry in the lattice structure to discuss the role of the DM interaction, whose direction is determined by the point-group symmetry and $\bm{q}$. 
The microscopic origin of the DM interaction is relativistic spin--orbit coupling. 
We neglect the symmetric anisotropic exchange interaction that arises from the higher-order contribution of the spin--orbit coupling for simplicity, although it also becomes the origin of the multiple-$Q$ states~\cite{Kato_PhysRevB.104.224405, yambe2023anisotropic}. 
The second term represents the Zeeman coupling through an external magnetic field, which tends to polarize the spins along the field direction $\bm{H}$. 

The ground-state magnetic instability in the model in Eq.~(\ref{eq: Ham1}) has been studied especially for the chiral point groups $O$ and $T$ to have $\bm{D}_{\bm{q}} \parallel \bm{q}$~\cite{Binz_PhysRevB.74.214408, Muhlbauer_2009skyrmion, Yi_PhysRevB.80.054416, Buhrandt_PhysRevB.88.195137, hayami2021field, Sabri_PhysRevB.107.024417}. 
In this case, the helical spiral state, whose spiral plane is perpendicular to the ordering vector $\bm{q}^*$, is stabilized at zero field. 
When the magnetic field is applied, the SkX, which is represented by a superposition of multiple spiral states along the symmetry-equivalent ordering vectors with $\bm{q}^*$, appears instead of the single-$Q$ helical spiral state. 
Although the characteristic ordering vector $\bm{q}^*$ is dependent on the magnitude and direction of $\bm{H}$, the spin configurations are well characterized by $\bm{q}^*$ and its symmetry-related ordering vectors once they are determined by $\bm{H}$. 
In other words, only a few characteristic ordering vectors in the Brillouin zone contribute to the energy in the ground state. 

Then, one can simplify the model in Eq.~(\ref{eq: Ham1}) by extracting the representative $\bm{q}^*$, which is transformed as 
\begin{align}
\label{eq: Ham2}
\mathcal{H}=&-\sum_{\nu}  \left[J \bm{S}_{\bm{Q}_\nu} \cdot  \bm{S}_{-\bm{Q}_\nu} 
+ i \bm{D}_\nu \cdot (\bm{S}_{\bm{Q}_\nu} \times  \bm{S}_{-\bm{Q}_\nu})\right] \nonumber \\
&-\sum_i \bm{H}\cdot \bm{S}_i, 
\end{align}
where we consider the specific ordering vectors $\bm{Q}_\nu \in \bm{q}^*$ giving the dominant contributions to the energy; $\nu$ is the label of the symmetry-equivalent wave vectors and the position of $\bm{Q}_\nu$ is fixed for simplicity. 
We set the same coupling constants $J$ and $D \equiv |\bm{D}_\nu|$ in the $\bm{Q}_\nu$ channel from the symmetry; we set $J=1$ as the energy unit of the model.

\begin{figure}[t!]
\begin{center}
\includegraphics[width=0.65\hsize]{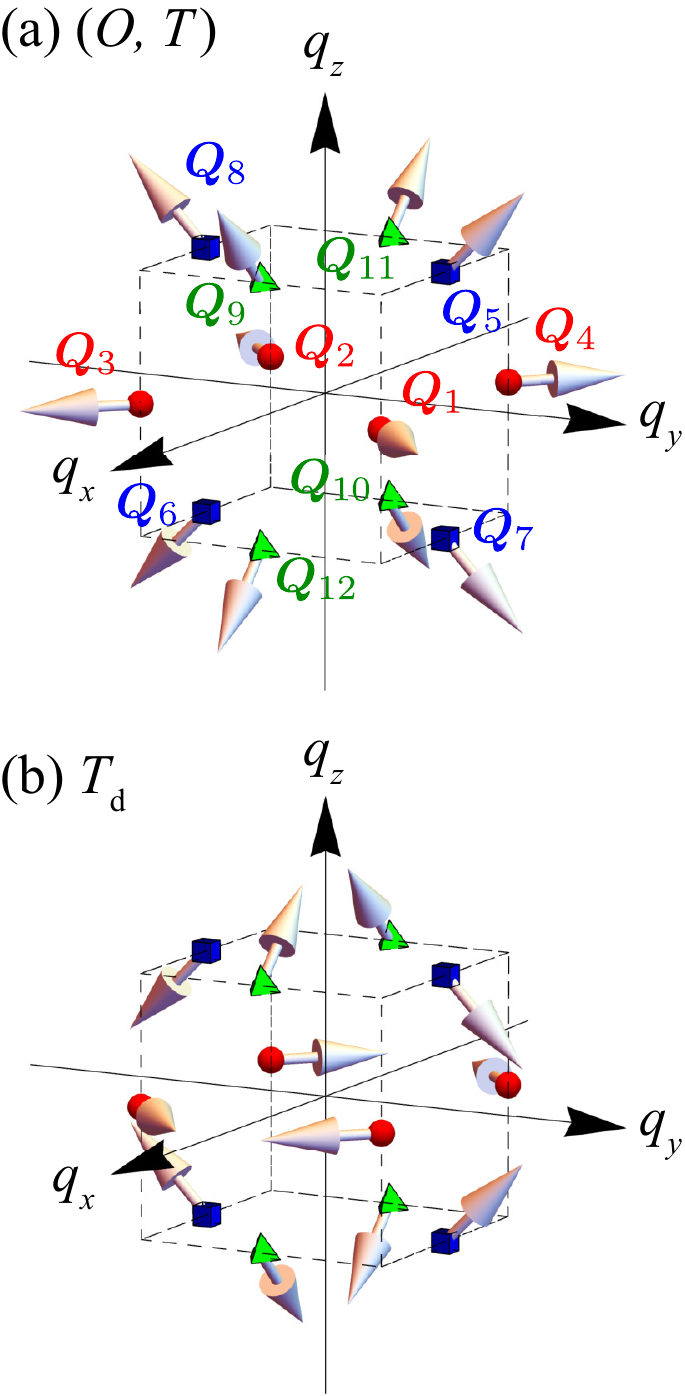} 
\caption{
\label{fig: Dvector} 
The DM vector at $\bm{Q}_\nu$ ($\nu=1$--{12}) in momentum space for (a) the point groups $(O,T)$ [Eq.~(\ref{eq: DM_O})] and (b) $T_{\rm d}$ [Eq.~(\ref{eq: DM_Td})]. 
}
\end{center}
\end{figure}

Meanwhile, the direction of $\bm{D}$ is determined by the symmetry in terms of $\bm{Q}_\nu$ and space group in the system~\cite{Yambe_PhysRevB.106.174437, yambe2023anisotropic}. 
We consider three noncentrosymmetric point (space) groups, $O$ ($P432$), $T$ ($P23$) and $T_{\rm d}$ ($P\bar{4}3m$), to discuss the similarities and differences in terms of the magnetic instability in the presence of the DM interaction. 
Since no DM interaction appears for $\bm{Q}_\nu \parallel [100]$ and $\bm{Q}_\nu \parallel [111]$ in the case of $T_{\rm d}$ ($P\bar{4}3m$)~\cite{yambe2023anisotropic}, we consider twelve ordering vectors lying on the $\langle 110 \rangle $ line, i.e., $\bm{Q}_\nu \parallel \langle 110 \rangle$ [Fig.~\ref{fig: Dvector}(a)]: 
$\bm{Q}_1=Q(1,1,0)$, 
$\bm{Q}_2=Q(-1,-1,0)$, 
$\bm{Q}_3=Q(1,-1,0)$, 
$\bm{Q}_4=Q(-1,1,0)$,
$\bm{Q}_5=Q(0,1,1)$, 
$\bm{Q}_6=Q(0,-1,-1)$, 
$\bm{Q}_7=Q(0,1,-1)$, 
$\bm{Q}_8=Q(0,-1,1)$,
$\bm{Q}_9=Q(1,0,1)$, 
$\bm{Q}_{10}=Q(-1,0,-1)$, 
$\bm{Q}_{11}=Q(-1,0,1)$, and 
$\bm{Q}_{12}=Q(1,0,-1)$, 
with $Q=\pi/3$.  
The difference between $(O,T)$ and $T_{\rm d}$ is found in the direction of $\bm{D}_\nu$ in each $\bm{Q}_\nu$: 
In the case of the point group $(O,T)$, $\bm{D}_\nu$ is given by 
\begin{align}
\label{eq: DM_O}
\bm{D}_1&= D(1,1,0), \nonumber \\
\bm{D}_3&= D(1,-1,0), \nonumber \\
\bm{D}_5&= D(0,1,1), \nonumber \\
\bm{D}_7&= D(0,1,-1), \nonumber \\
\bm{D}_9&= D(1,0,1), \nonumber \\
\bm{D}_{11}&= D(-1,0,1), 
\end{align}
and $\bm{D}_{2\zeta}=-\bm{D}_{2\zeta-1}$ for $\zeta=1$--$6$. 
This type of DM interaction is referred to as the chiral-type DM interaction with $\bm{D}_\nu \parallel \bm{Q}_\nu$. 
On the other hand, $\bm{D}_\nu$ for the point group $T_{\rm d}$ is given by 
\begin{align}
\label{eq: DM_Td}
\bm{D}_1&= D(1,-1,0), \nonumber \\
\bm{D}_3&= D(1,1,0), \nonumber  \\
\bm{D}_5&= D(0,1,-1), \nonumber \\
\bm{D}_7&= D(0,1,1), \nonumber \\
\bm{D}_9&= D(-1,0,1), \nonumber \\
\bm{D}_{11}&= D(1,0,1), 
\end{align}
and $\bm{D}_{2\zeta}=-\bm{D}_{2\zeta-1}$ for $\zeta=1$--$6$. 
This type of DM interaction is referred to as the rank-3 polar-type DM interaction with $\bm{D}_\nu \perp \bm{Q}_\nu$~\cite{comment_polarDM, Ado_PhysRevB.101.161403}. 
The directions of the DM vectors at $\bm{Q}_\nu$ for $(O,T)$ and $T_{\rm d}$ are schematically shown in Figs.~\ref{fig: Dvector}(a) and \ref{fig: Dvector}(b), respectively.

For the effective spin model in Eq.~(\ref{eq: Ham2}) with the DM interaction in Eq.~(\ref{eq: DM_O}) or (\ref{eq: DM_Td}), we investigate a low-temperature phase diagram while changing $D$ and $H \equiv |\bm{H}|$ by performing the simulated annealing following the manner in Ref.~\cite{Hayami_PhysRevB.95.224424} as follows. 
Starting from a random spin configuration at high temperatures $T_0=$1--10, we gradually reduce the temperature ($T$) by the rate $\alpha=0.999999$ to the target low temperature $T_f=0.001$ in each Monte Carlo sweep; the local updates are performed in real space based on the standard Metropolis algorithm and we sample homogeneously on a unit sphere for the updates. 
At the target temperature, we perform $10^5$--$10^6$ Monte Carlo sweeps for measurements after equilibration. 
The simulations are independently performed for a given set of $(D, H)$ in Sec.~\ref{sec: Comparison} and $(\kappa, H)$ in Sec.~\ref{sec: Uniaxial strain effect}. 
When the model parameters are close to those at the phase boundaries, we also start the simulations from the spin configurations obtained at low temperatures to avoid the local minima. 
The system size is taken as $N=12^3$ spins in order to be commensurate with the lattice structure; the position vector at each site is given by $\bm{r}_i=(x_i, y_i, z_i)$ where $x_i$, $y_i$, and $z_i$ are integers from 0 to 11. 
The boundary condition is the periodic boundary. 

To identify the magnetic phases obtained by the simulated annealing, we compute the uniform magnetization 
\begin{align}
\label{eq: magnetization}
M^\eta=\frac{1}{N} \sum_i S_i^\eta, 
\end{align}
for $\eta=x,y,z$ and the spin structure factor divided by $N$, which corresponds to the magnetic moments, with wave vector $\bm{Q}_\nu$
\begin{align}
(m_{\bm{Q}_\nu})^2&=(m^x_{\bm{Q}_\nu})^2+(m^y_{\bm{Q}_\nu})^2+(m^z_{\bm{Q}_\nu})^2,  \\
(m^\eta_{\bm{Q}_\nu})^2&=\frac{1}{N^2}\sum_{ij}S_i^\eta S_j^\eta e^{i \bm{Q}_\nu \cdot (\bm{r}_i -\bm{r}_j)}.
\end{align}
It is noted $m_{\bm{Q}_{2\zeta-1}}=m_{\bm{Q}_{2\zeta}}$ for $\zeta=1$--$6$. 
We also evaluate the spin scalar chirality on the $\eta\eta'=xy, yz, zx$ plane as
\begin{align}
\chi^{\eta\eta'}=\frac{1}{N} \sum_{i \delta_\eta \delta_{\eta'}}\delta_{\eta}\delta_{\eta'} \bm{S}_i \cdot (\bm{S}_{i+\delta_\eta \hat{\bm{x}}_\eta} \times \bm{S}_{i+\delta_{\eta'} \hat{\bm{x}}_{\eta'}}),
\end{align}
where $\delta_{\eta}, \delta_{\eta'}=\pm 1$ and $\hat{\bm{x}}_\eta$ is the $\eta$-directional unit vector. 

\section{Results}
\label{sec: Results}

\subsection{Comparison of phase diagrams between point groups $(O,T)$ and $T_{\rm d}$}
\label{sec: Comparison}

\begin{figure*}[t!]
\begin{center}
\includegraphics[width=1.0\hsize]{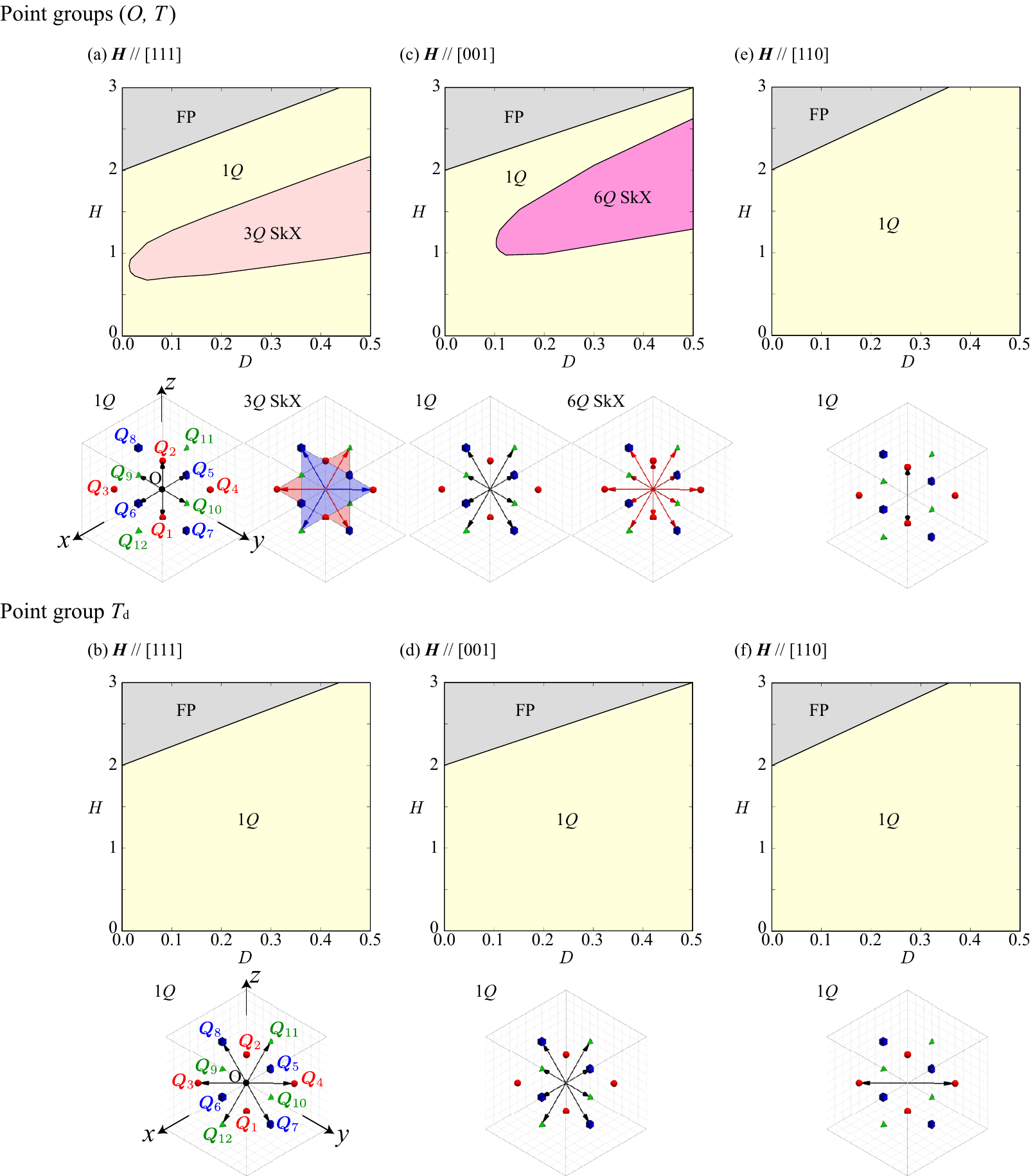} 
\caption{
\label{fig: PD} 
Phase diagrams of the model in Eq.~(\ref{eq: Ham2}) under (a), (c), (e) the point groups $(O,T)$ and (b), (d), (f) $T_{\rm d}$ in the $D$--$H$ plane obtained by the simulated annealing. 
The magnetic field directions are taken along (a,b) the [111] direction, (c,d) [001] direction, and (e,f) [110] direction. 
1$Q$, 3$Q$ SkX, 6$Q$ SkX, and FP represent the single-$Q$ state, triple-$Q$ skyrmion crystal, sextuple-$Q$ skyrmion crystal, and the fully-polarized state, respectively. 
The lower panels stand for the ordering wave vectors among $\bm{Q}_1$--$\bm{Q}_{12}$.
}
\end{center}
\end{figure*}

We construct the phase diagram of the model in Eq.~(\ref{eq: Ham2}) with an emphasis on the similarities and differences of the multiple-$Q$ instability between the point groups $(O,T)$ and $T_{\rm d}$, which are characterized by the different DM vectors.  
We show the six phase diagrams in the $D$--$H$ plane in Fig.~\ref{fig: PD}, which are obtained for the three different orientations of $\bm{H}$ and two point groups $(O,T)$ and $T_{\rm d}$.
We discuss the results for $\bm{H} \parallel [111]$ in Sec.~\ref{sec: [111]}, $\bm{H} \parallel [001]$ in Sec.~\ref{sec: [001]}, and $\bm{H}\parallel [110]$ in Sec.~\ref{sec: [110]}.

\subsubsection{[111] magnetic field}
\label{sec: [111]}

Figure~\ref{fig: PD}(a) shows the $D$--$H$ phase diagram for the model in Eq.~(\ref{eq: Ham2}) with the DM interaction in Eq.~(\ref{eq: DM_O}) under the [111] magnetic field. 
The phase diagram consists of three phases: the single-$Q$ (1$Q$) state, the triple-$Q$ SkX (3$Q$ SkX) and the fully-polarized (FP) state. 
The 1$Q$ state at zero field is characterized by the proper-screw spiral state in order to gain the energy by the DM interaction; the spiral planes are perpendicular to $\bm{Q}_\nu$.  
The ordering vector is chosen out of $\bm{Q}_1$--$\bm{Q}_{12}$ depending on initial spin configurations; the spiral states with any of $\bm{Q}_\eta$ have the same energy with each other. 

When the magnetic field is turned on, the ordering vector parallel to the $[111]$ axis as much as possible, i.e., $\bm{Q}_1$, $\bm{Q}_5$, or $\bm{Q}_9$ ($\bm{Q}_2$, $\bm{Q}_6$, or $\bm{Q}_{10}$), is chosen as the lowest-energy state. 
This is because the proper-screw spiral configuration for these ordering vectors has more perpendicular spin components to $\bm{H}$ than that for the other ordering vectors, which results in the energy gain by the Zeeman coupling. 

\begin{figure}[t!]
\begin{center}
\includegraphics[width=1.0\hsize]{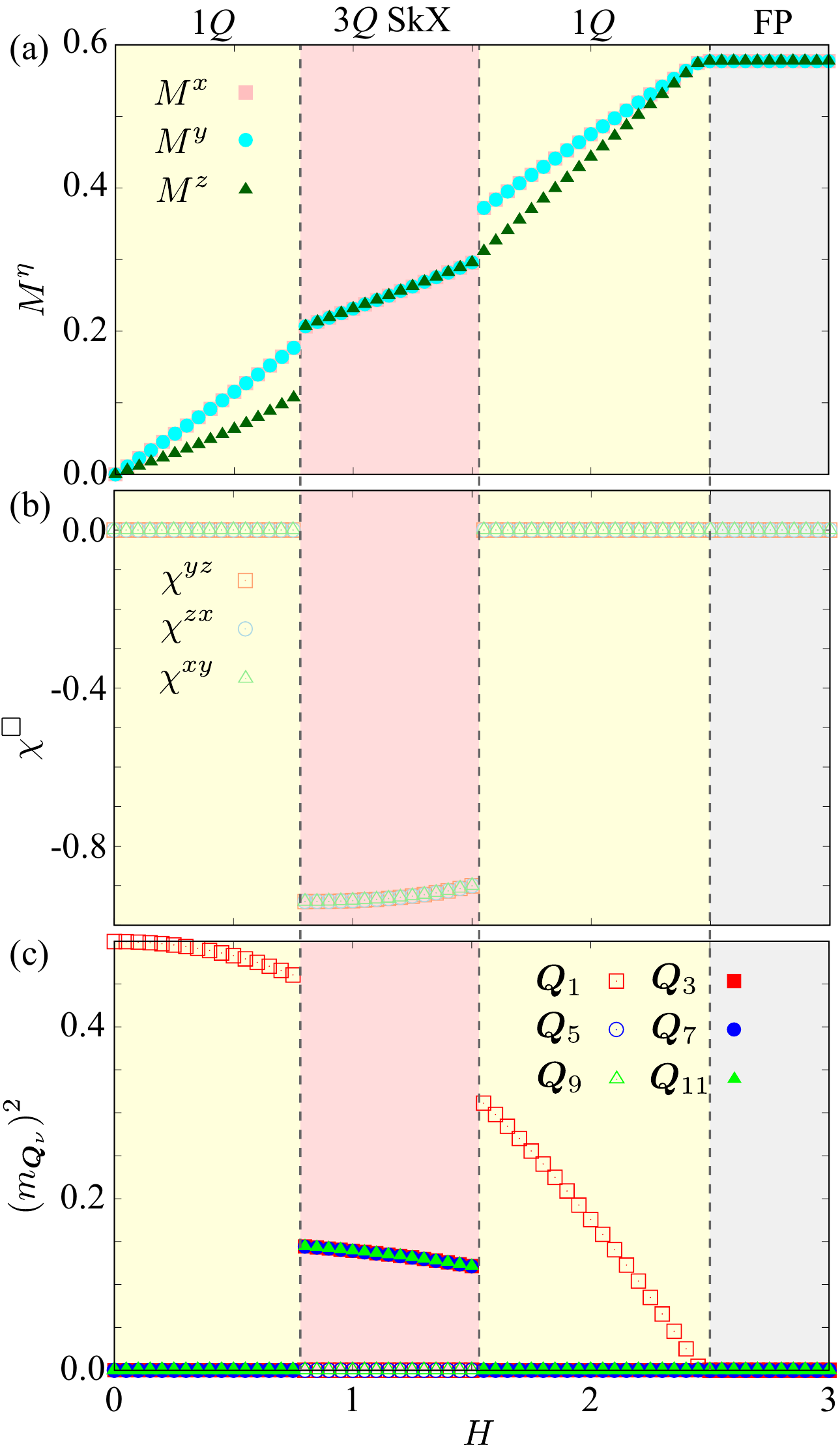} 
\caption{
\label{fig: mag_111} 
$H$ dependence of (a) the magnetization $M^\eta$ for $\eta=x,y,z$, (b) the scalar chirality $\chi^{\square}$ for $\square=yz, zx, xy$, and (c) $(m_{\bm{Q}_{\nu}})^2$ for $\nu=1,3,5,7,9,11$ at $D=0.2$ for the model in Eq.~(\ref{eq: Ham2}) under the [111] field. 
The vertical dashed lines show the phase boundaries. 
}
\end{center}
\end{figure}

\begin{figure}[t!]
\begin{center}
\includegraphics[width=1.0\hsize]{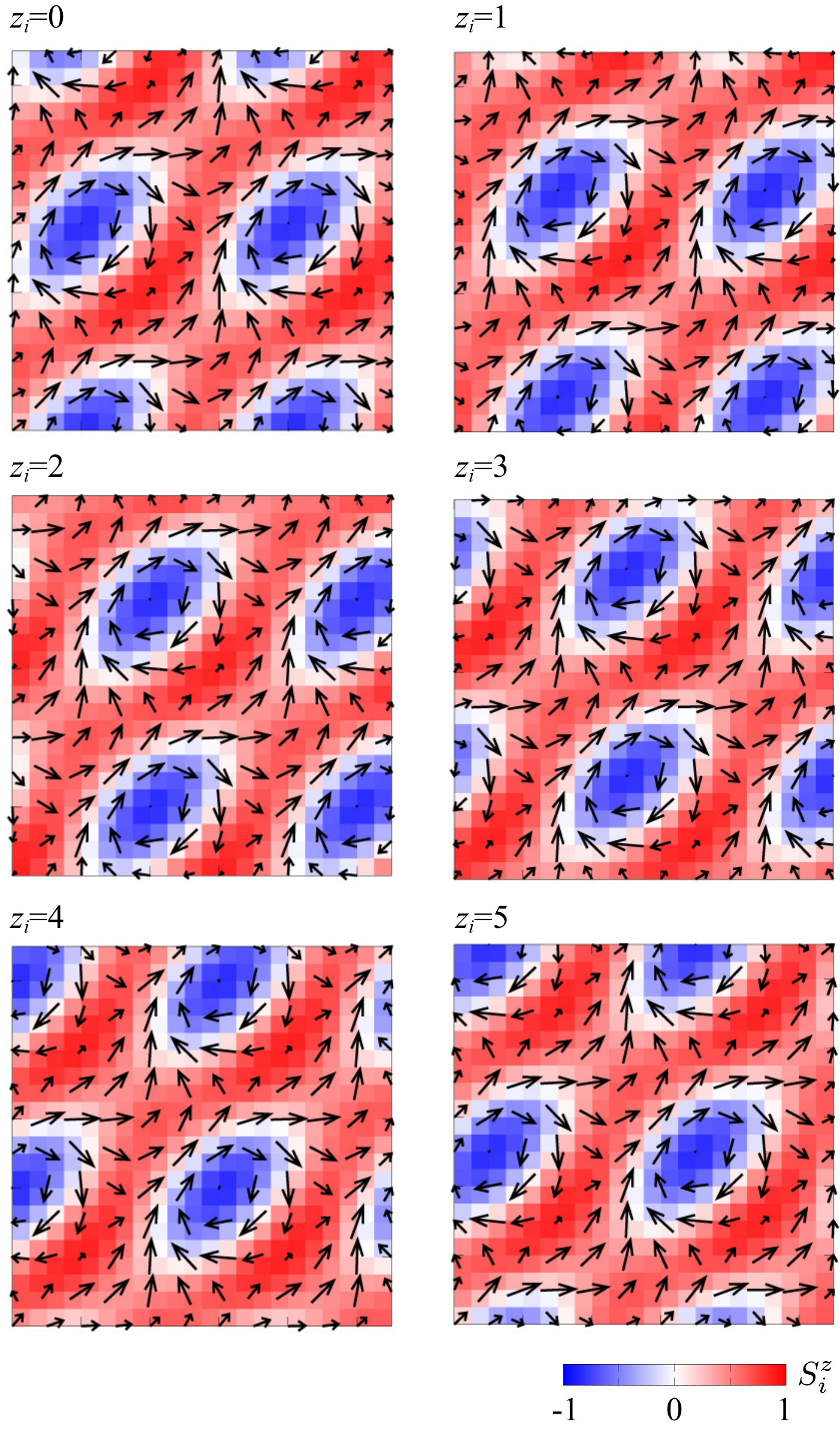} 
\caption{
\label{fig: spin} 
Real-space spin configuration of the 3$Q$ SkX at $D=0.2$ and $H=1$ for $\bm{H} \parallel [111]$ on the $xy$ plane at $z_i=0$--$5$. 
The arrows represent the in-plane spin components, while the color represents the out-of-plane spin component.  
The spin configuration is drawn by the single snapshot at the lowest temperature.
}
\end{center}
\end{figure}

In the intermediate-field region, the 1$Q$ state is replaced by the 3$Q$ SkX with the jumps of magnetization $M^{\eta}$ for $\eta=x,y,z$, as shown in Fig.~\ref{fig: mag_111}(a). 
The region of the 3$Q$ SkX becomes wider for larger $D$. 
The spin configuration of the 3$Q$ SkX on the $xy$ plane for different $z$-coordinate is shown in Fig.~\ref{fig: spin}. 
The skyrmion core located at $S_i^z = -1$ is elliptically distorted and it forms the square lattice in each $xy$ plane, which indicates the emergence of nonzero spin scalar chirality in the $xy$ plane, i.e., $\chi^{xy} \neq 0$, as shown in Fig.~\ref{fig: mag_111}(b).
Meanwhile, the 3$Q$ SkX is characterized by a triple-$Q$ superposition of the spiral states at $\bm{Q}_3$, $\bm{Q}_7$, and $\bm{Q}_{11}$ ($\bm{Q}_4$, $\bm{Q}_8$, and $\bm{Q}_{12}$), which are connected by threefold rotational symmetry around the [111] axis, with the same intensity shown in Fig.~\ref{fig: mag_111}(c): $m^x_{\bm{Q}_3}=m^y_{\bm{Q}_3}=m^y_{\bm{Q}_7}=m^z_{\bm{Q}_7}=m^z_{\bm{Q}_{11}}=m^x_{\bm{Q}_{11}}$ and $m^z_{\bm{Q}_3}=m^x_{\bm{Q}_7}=m^y_{\bm{Q}_{11}}$. 
The choice of these three ordering vectors is owing to the relation $\bm{Q}_3 + \bm{Q}_7 + \bm{Q}_{11} = \bm{0}$ ($\bm{Q}_4 + \bm{Q}_8 + \bm{Q}_{12} = \bm{0}$), which tends to avoid the higher-harmonics contribution to hamper the formation of multiple-$Q$ states. 
In addition, the triple-$Q$ ordering vectors are selected so as to be perpendicular to the field direction for the same reason. 
Reflecting the presence of the threefold symmetry around the [111] direction, the triangular-lattice alignment of the skyrmion can be seen onto the (111) plane, i.e., $\chi^{xy}=\chi^{yz}=\chi^{zx}$. 

With a further increase of $H$, the 3$Q$ SkX turns into the 1$Q$ state, whose transition is of first-order, as shown in Fig.~\ref{fig: mag_111}. 
Finally, the 1$Q$ state continuously changes into the FP state. 
Such transitions between the 1$Q$ state, 3$Q$ SkX, and FP state are also found for a different choice of the ordering vectors~\cite{hayami2021field}.

In contrast, the model with the DM interaction under the point group $T_{\rm d}$ in Eq.~(\ref{eq: DM_Td}) exhibits a qualitatively different phase diagram, as shown in Fig.~\ref{fig: PD}(b); no 3$Q$ SkX appears for nonzero $D$ and $H$. 
In addition, the 1$Q$ state is characterized by different ordering vectors from the case of $(O,T)$ in the lower panel of Fig.~\ref{fig: PD}(b). 
The 1$Q$ state at zero field is characterized by the cycloidal spiral state in order to gain the energy by the DM interaction; the spiral planes for $\bm{Q}_1$-$\bm{Q}_4$, $\bm{Q}_5$-$\bm{Q}_8$, and $\bm{Q}_9$-$\bm{Q}_{12}$ are parallel to $\bm{\hat{z}}$, $\bm{\hat{x}}$, and $\bm{\hat{y}}$, respectively, as well as the ordering wave vector $\bm{Q}_\nu$.  

When the magnetic field is turned on, the ordering vector perpendicular to the $[111]$ axis, i.e., $\bm{Q}_3$, $\bm{Q}_7$, or $\bm{Q}_{11}$ ($\bm{Q}_4$, $\bm{Q}_8$, or $\bm{Q}_{12}$), is chosen as the lowest-energy state. 
This is because the spiral plane under the point group $T_{\rm d}$ tends to be parallel to $\bm{Q}_\nu$ in order to gain energy by the DM interaction.  
In the end, the spiral state with the ordering vector perpendicular to the [111] axis gains both energies by the DM interaction and Zeeman coupling. 
When $H$ increases, this state turns into the FP state.

\subsubsection{[001] magnetic field}
\label{sec: [001]}

\begin{figure}[t!]
\begin{center}
\includegraphics[width=1.0\hsize]{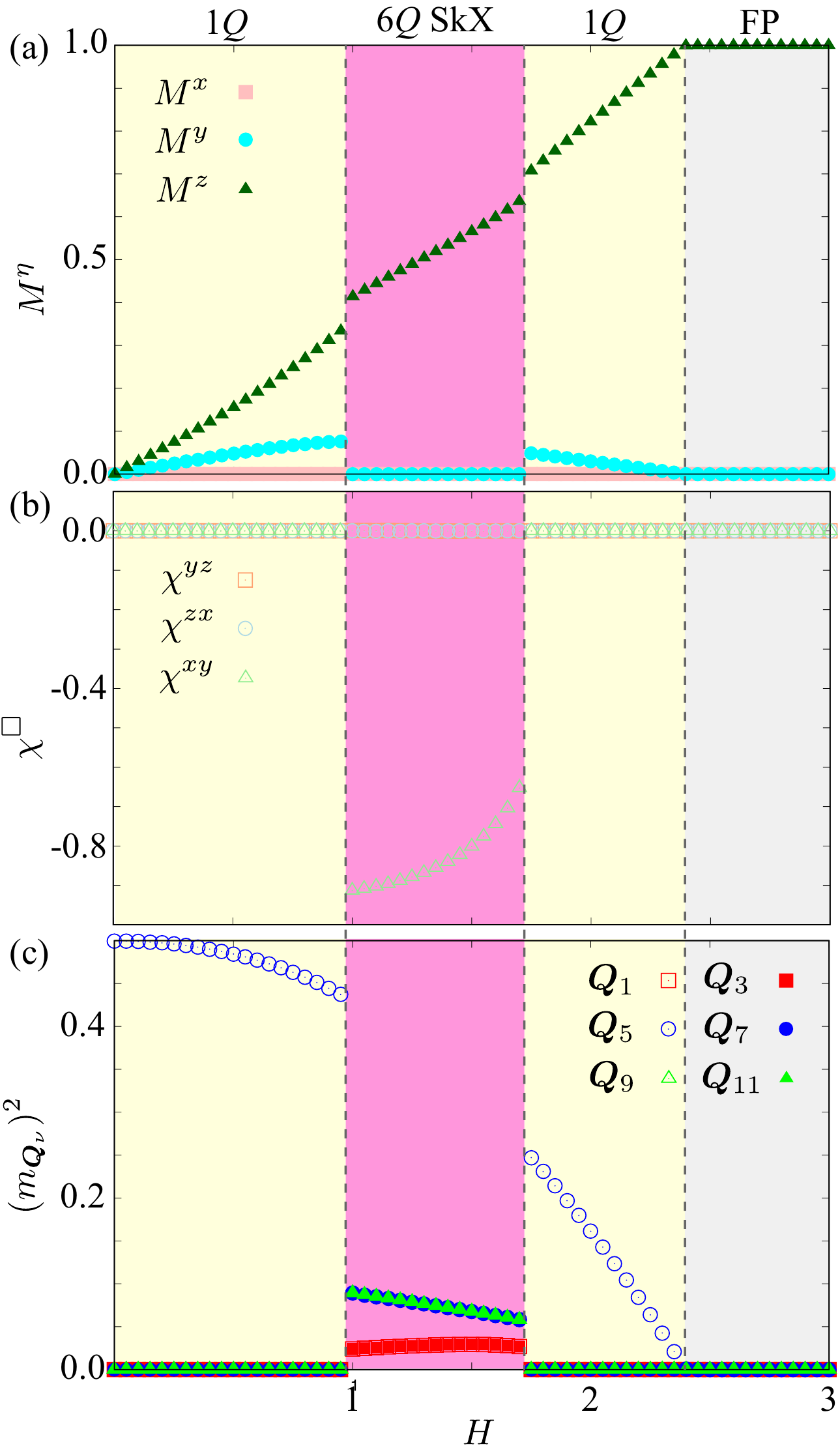} 
\caption{
\label{fig: mag_001} 
$H$ dependence of (a) the magnetization $M^\eta$ for $\eta=x,y,z$, (b) the scalar chirality $\chi^{\square}$ for $\square=yz, zx, xy$, and (c) $(m_{\bm{Q}_{\nu}})^2$ for $\nu=1,3,5,7,9,11$ at $D=0.2$ for the model in Eq.~(\ref{eq: Ham2}) under the [001] field. 
The vertical dashed lines show the phase boundaries. 
}
\end{center}
\end{figure}

\begin{figure}[t!]
\begin{center}
\includegraphics[width=1.0\hsize]{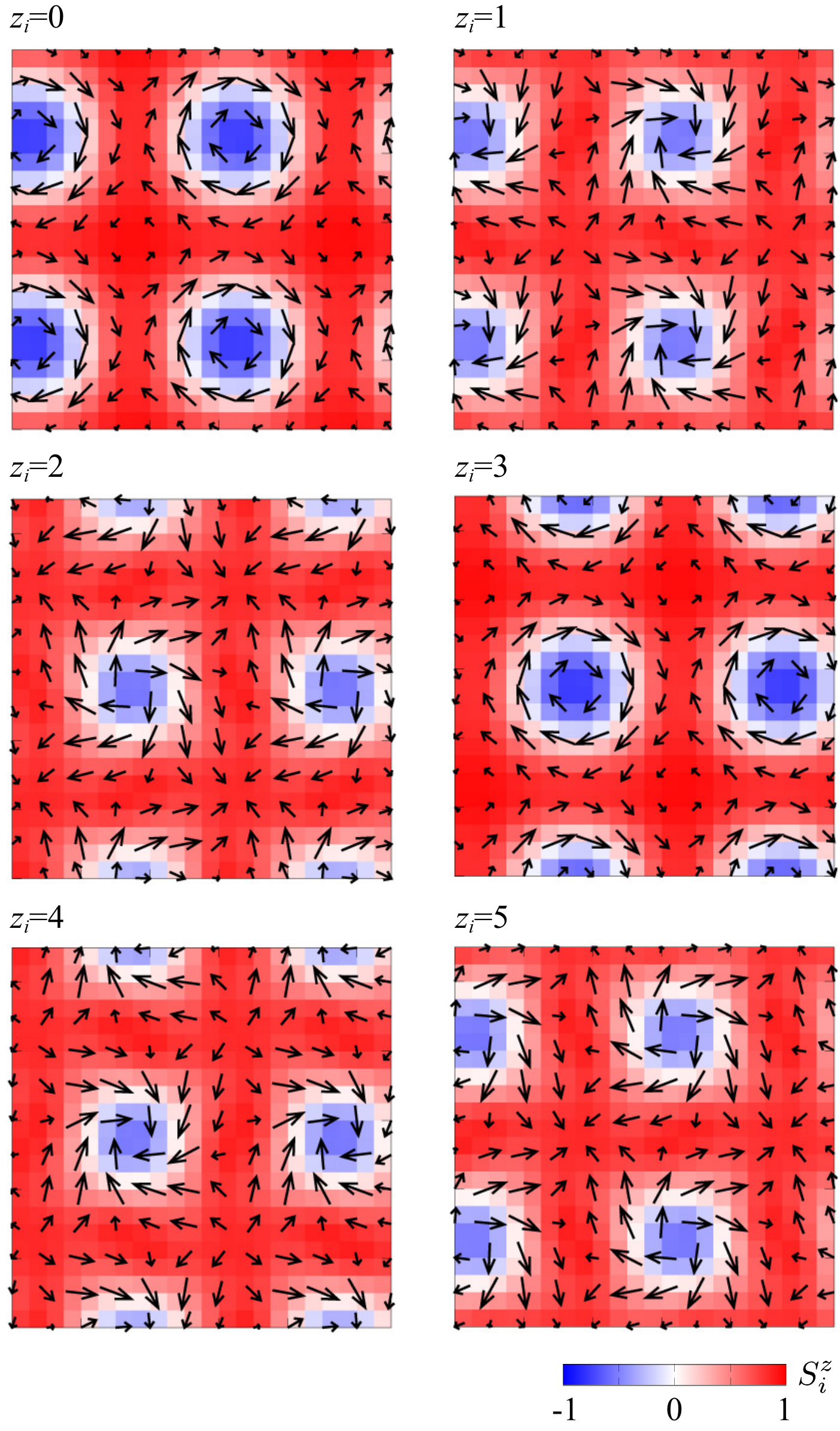} 
\caption{
\label{fig: spin_skx6} 
Real-space spin configuration of the 6$Q$ SkX at $D=0.2$ and $H=1.2$ for $\bm{H} \parallel [001]$ on the $xy$ plane at $z_i=0$--$5$. 
The arrows represent the in-plane spin components, while the color represents the out-of-plane spin component.  
The spin configuration is drawn by the single snapshot at the lowest temperature.
}
\end{center}
\end{figure}

Next, we consider the situation where the magnetic field is applied along the [001] direction. 
The phase diagrams for the point group $(O,T)$ and $T_{\rm d}$ are shown in Figs.~\ref{fig: PD}(c) and \ref{fig: PD}(d), respectively. 
The tendency of the multiple-$Q$ instability is similar to that in the [111] field; the SkX is stabilized for the case of $(O,T)$, while no SkX appears for the case of $T_{\rm d}$.

As shown in the lower panel of Fig.~\ref{fig: PD}(c), the ordering vectors for the 1$Q$ state under $(O,T)$ is given by $\bm{Q}_5$, $\bm{Q}_7$, $\bm{Q}_9$, or $\bm{Q}_{11}$ ($\bm{Q}_6$, $\bm{Q}_8$, $\bm{Q}_{10}$, or $\bm{Q}_{12}$).
Similar to the [111] field, the ordering vectors with a large parallel component to the field are chosen, as shown in the lower panel of Fig.~\ref{fig: PD}(c). 
This is because the proper-screw spiral at these wave vectors has more spin components perpendicular to the [001] field than $\bm{Q}_1$--$\bm{Q}_4$. 
On the other hand, the sextuple-$Q$ SkX (6$Q$ SkX) appears for $D \gtrsim 0.1$ with the increase of the magnetic field from the region of the 1$Q$ state in Fig.~\ref{fig: PD}(c); there are sudden increases of $M^\eta$ and $\chi^{xy}$, as shown in Figs.~\ref{fig: mag_001}(a) and \ref{fig: mag_001}(b), respectively. 
The spin configuration of the 6$Q$ SkX is presented in Fig.~\ref{fig: spin_skx6} at $D=0.2$ and $H=1.2$. 
In contrast to the elliptical skyrmion core in the 3$Q$ SkX in Fig.~\ref{fig: spin}, the skyrmion core in the 6$Q$ SkX has a circular shape.
Meanwhile, the skyrmion core in the 6$Q$ SkX forms the square lattice as well as the 3$Q$ SkX. 
The different skyrmion core position in different $z$-coordinate indicates the modulation along the 
$z$ direction. 
Indeed, there are magnetic moments at $\bm{Q}_5$, $\bm{Q}_7$, $\bm{Q}_9$, and $\bm{Q}_{11}$ with the same intensity as well as those at $\bm{Q}_1$ and $\bm{Q}_3$, as shown in Fig.~\ref{fig: mag_001}(c), which indicates the translation of the skyrmion core in the $xy$ plane at different $z$-coordinate~\cite{Shimizu_PhysRevB.105.224405}.
Moreover, these additional contributions from $\bm{Q}_5$, $\bm{Q}_7$, $\bm{Q}_9$, and $\bm{Q}_{11}$ modulate the alignment of the SkX core from the two-dimensional SkX with $m_{\bm{Q}_1}=m_{\bm{Q}_3}\neq 0$ and $m_{\bm{Q}_5} = m_{\bm{Q}_7}=m_{\bm{Q}_9}=m_{\bm{Q}_{11}}=0$: The former has the nearest-neighbor skyrmion core along the $\langle 100 \rangle$ direction, while the latter has that along the $\langle 110 \rangle $ direction. 
In each layer, the helicity of the SkX is different but its vorticity is the same as each other so as to have a nonzero skyrmion number in the whole system. 
It is noted that the magnetic structure has fourfold rotational symmetry around the $z$ axis. 

The appearance of the 6$Q$ SkX rather than the double-$Q$ SkX to have only nonzero $m_{\bm{Q}_1}$ and $m_{\bm{Q}_3}$ but $m_{\bm{Q}_5}=m_{\bm{Q}_7}=m_{\bm{Q}_9}=m_{\bm{Q}_{11}}=0$ is a consequence of the three-dimensional system so that the $z$-directional modulation can reduce the energy. 
Indeed, the relations of ${\bm{Q}_1}=-\bm{Q}_5-\bm{Q}_{10}$ and ${\bm{Q}_3}=-\bm{Q}_7-\bm{Q}_{11}$ indicate the appearance of effective couplings under the magnetic field, $(\bm{m}_{\bm{0}}\cdot \bm{m}_{\bm{Q}_1})(\bm{m}_{\bm{Q}_5}\cdot \bm{m}_{\bm{Q}_{10}})$ and $(\bm{m}_{\bm{0}}\cdot \bm{m}_{\bm{Q}_3})(\bm{m}_{\bm{Q}_7}\cdot \bm{m}_{\bm{Q}_{11}})$, which lowers the energy compared to the double-$Q$ SkX; one can find that the relations of ${\bm{Q}_1}+\bm{Q}_5+\bm{Q}_{10}=\bm{0}$ and ${\bm{Q}_3}+\bm{Q}_7+\bm{Q}_{11}=\bm{0}$ are important in the effective coupling owing to the momentum conservation.
Thus, the 6$Q$ SkX is a characteristic phase that arises from the interplay among the Heisenberg interaction, the DM interaction, and the magnetic field at three-dimensional ordering vectors. 
Similar multiple-$Q$ states characterized by more than four wave vectors have been also clarified for three-dimensional similar models under noncentrosymmetric lattice structures~\cite{Kato_PhysRevB.105.174413, Kato_PhysRevB.107.094437}.

Meanwhile, the phase diagram consists of the 1$Q$ and FP states in the case of $T_{\rm d}$. 
The 1$Q$ state under $T_{\rm d}$ is characterized by the ordering vector at $\bm{Q}_5$, $\bm{Q}_7$, $\bm{Q}_9$, or $\bm{Q}_{11}$ ($\bm{Q}_6$, $\bm{Q}_8$, $\bm{Q}_{10}$, or $\bm{Q}_{12}$), as shown in Fig.~\ref{fig: PD}(d).
This is because the cycloidal spiral at these wave vectors has more spin components perpendicular to the [001] field than that at $\bm{Q}_1$--$\bm{Q}_4$.

\subsubsection{[110] magnetic field}
\label{sec: [110]}

The phase diagrams under the [110] magnetic field for the point groups $(O,T)$ and $T_{\rm d}$ are shown in Figs.~\ref{fig: PD}(e) and \ref{fig: PD}(f), respectively. 
In both cases, no multiple-$Q$ instability occurs in the phase diagram; the 1$Q$ state with $\bm{Q}_1$ is stabilized for $(O,T)$ and that with $\bm{Q}_3$ is stabilized for $T_{\rm d}$; the conical spiral state, whose spiral plane is perpendicular to the magnetic field, is realized in both cases in order to gain the energy by the Zeeman interaction under the local spin-length constraint $|\bm{S}_i|=1$.  
Thus, the energy of the 1$Q$ state is always lower than that of the multiple-$Q$ state, the latter of which usually costs the energy by the higher-harmonic wave-vector contributions.

\subsection{Uniaxial strain effect for point group $T_{\rm d}$}
\label{sec: Uniaxial strain effect}

We have so far considered the multiple-$Q$ instability under the point groups $(O,T)$ and $T_{\rm d}$. 
In contrast to the system under the point group $(O,T)$ with the chiral-type DM interaction, the system under $T_{\rm d}$ with the rank-3 polar-type DM interaction does not lead to the multiple-$Q$ instability, which is consistent with the analysis based on Lifshitz invariants in the free energy~\cite{dzyaloshinskii1964theory,kataoka1981helical, Bogdanov89, Bogdanov94}. 
To explore the possibility of the multiple-$Q$ states under the $T_{\rm d}$ system, we consider the effect of uniaxial strain along the $z$ direction, which reduces the point group symmetry from $T_{\rm d}$ to $D_{\rm 2d}$, since the two-dimensional anisotropy tends to stabilize the multiple-$Q$ states including the SkX~\cite{tonomura2012real, seki2012observation}. 

We express the effect of the uniaxial anisotropy by the different interaction amplitudes for the $xy$ and $z$ directions in real space from the symmetry viewpoint; the tetragonal symmetry under the uniaxial strain leads to inequivalence of interactions at $xy$-plane and out-of-plane wave vectors in momentum space. 
The model Hamiltonian is given by 
\begin{align}
\label{eq: Ham_D2d}
\mathcal{H}=&-\sum_{\nu'}  \left[J \bm{S}_{\bm{Q}_{\nu'}} \cdot  \bm{S}_{-\bm{Q}_{\nu'}} 
+ i \bm{D}_{\nu'} \cdot (\bm{S}_{\bm{Q}_{\nu'}} \times  \bm{S}_{-\bm{Q}_{\nu'}})\right] \nonumber \\
&-\kappa\sum_{\nu''}  \left[J \bm{S}_{\bm{Q}_{\nu''}} \cdot  \bm{S}_{-\bm{Q}_{\nu''}} 
+ i \bm{D}_{\nu''} \cdot (\bm{S}_{\bm{Q}_{\nu''}} \times  \bm{S}_{-\bm{Q}_{\nu''}})\right] \nonumber \\
&-\sum_i \bm{H}\cdot \bm{S}_i, 
\end{align}
where $\nu'=1$--$4$ and $\nu''=5$--$12$. 
$\kappa<1$ represents the ratio of the interactions at $\bm{Q}_1$--$\bm{Q}_4$ and those at $\bm{Q}_5$--$\bm{Q}_{12}$. 
We fix $D=0.2$ and $\bm{H}=(0,0,H)$ in the following calculations.

\begin{figure}[t!]
\begin{center}
\includegraphics[width=1.0\hsize]{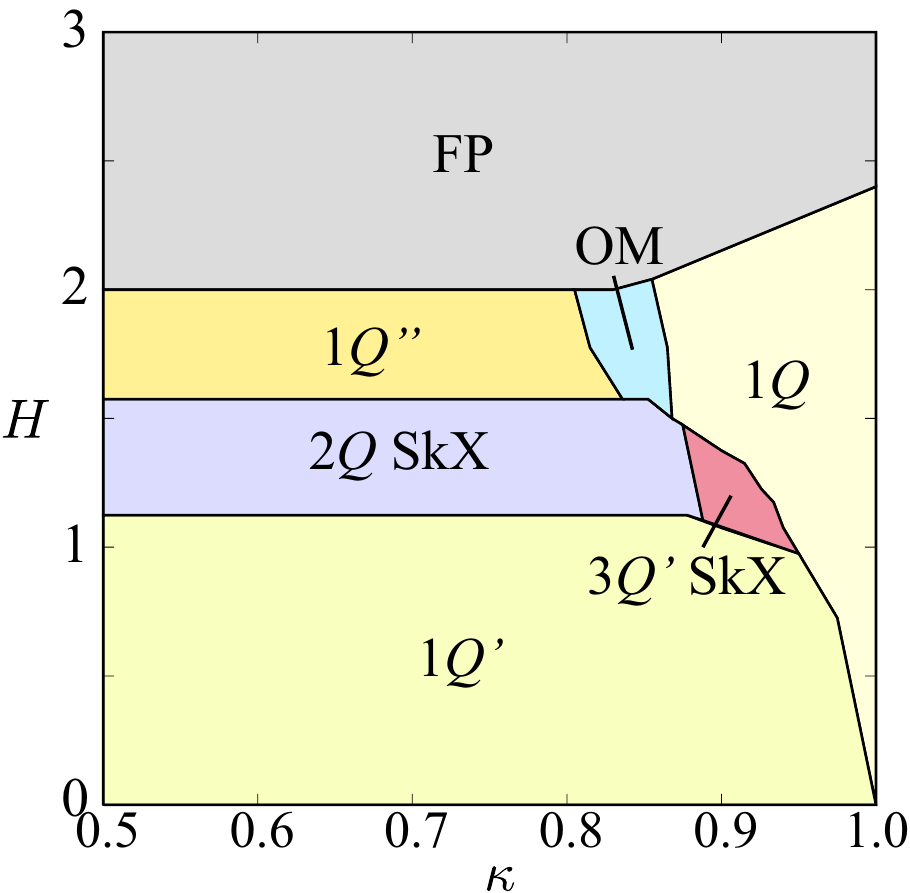} 
\caption{
\label{fig: PD2} 
$\kappa$--$H$ phase diagram of the model in Eq.~(\ref{eq: Ham_D2d}) at $D=0.2$ for $\bm{H}\parallel [001]$ under the point group $D_{\rm 2d}$. 
1$Q'$ and 1$Q''$ represent the different single-$Q$ states from the 1$Q$ state. 
2$Q$ SkX, 3$Q'$ SkX, and OM represent the double-$Q$ skyrmion crystal, different triple-$Q$ skyrmion crystal from 3$Q$ SkX in Fig.~\ref{fig: PD}, and other magnetic states, respectively. 
}
\end{center}
\end{figure}

\begin{figure}[t!]
\begin{center}
\includegraphics[width=1.0\hsize]{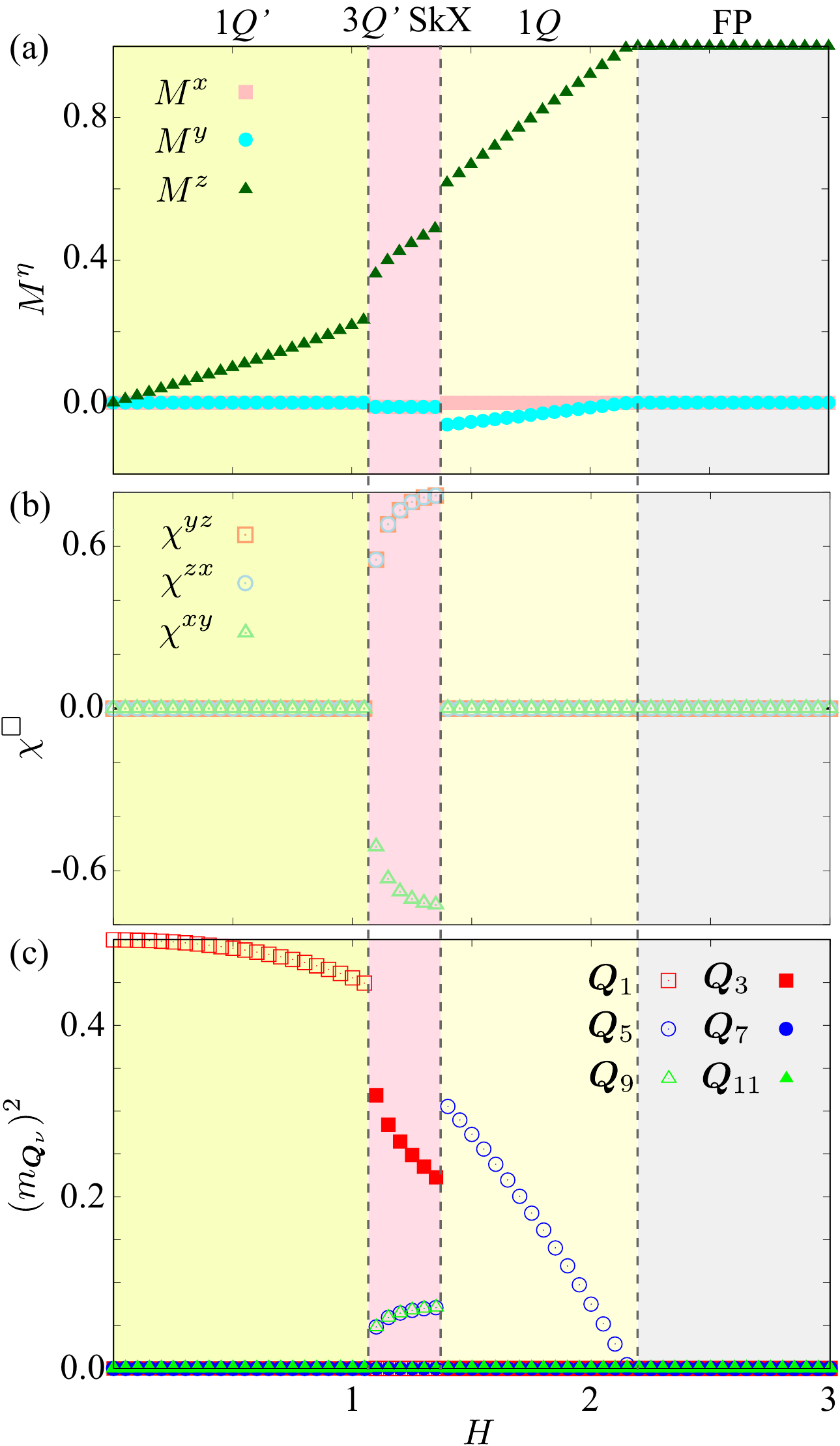} 
\caption{
\label{fig: mag_001_dist_1} 
$H$ dependence of (a) the magnetization $M^\eta$ for $\eta=x,y,z$, (b) the scalar chirality $\chi^{\square}$ for $\square=yz, zx, xy$, and (c) $(m_{\bm{Q}_{\nu}})^2$ for $\nu=1,3,5,7,9,11$ at $\kappa=0.9$ for the model in Eq.~(\ref{eq: Ham_D2d}) under the [001] field. 
The vertical dashed lines show the phase boundaries. 
}
\end{center}
\end{figure}

\begin{figure}[t!]
\begin{center}
\includegraphics[width=1.0\hsize]{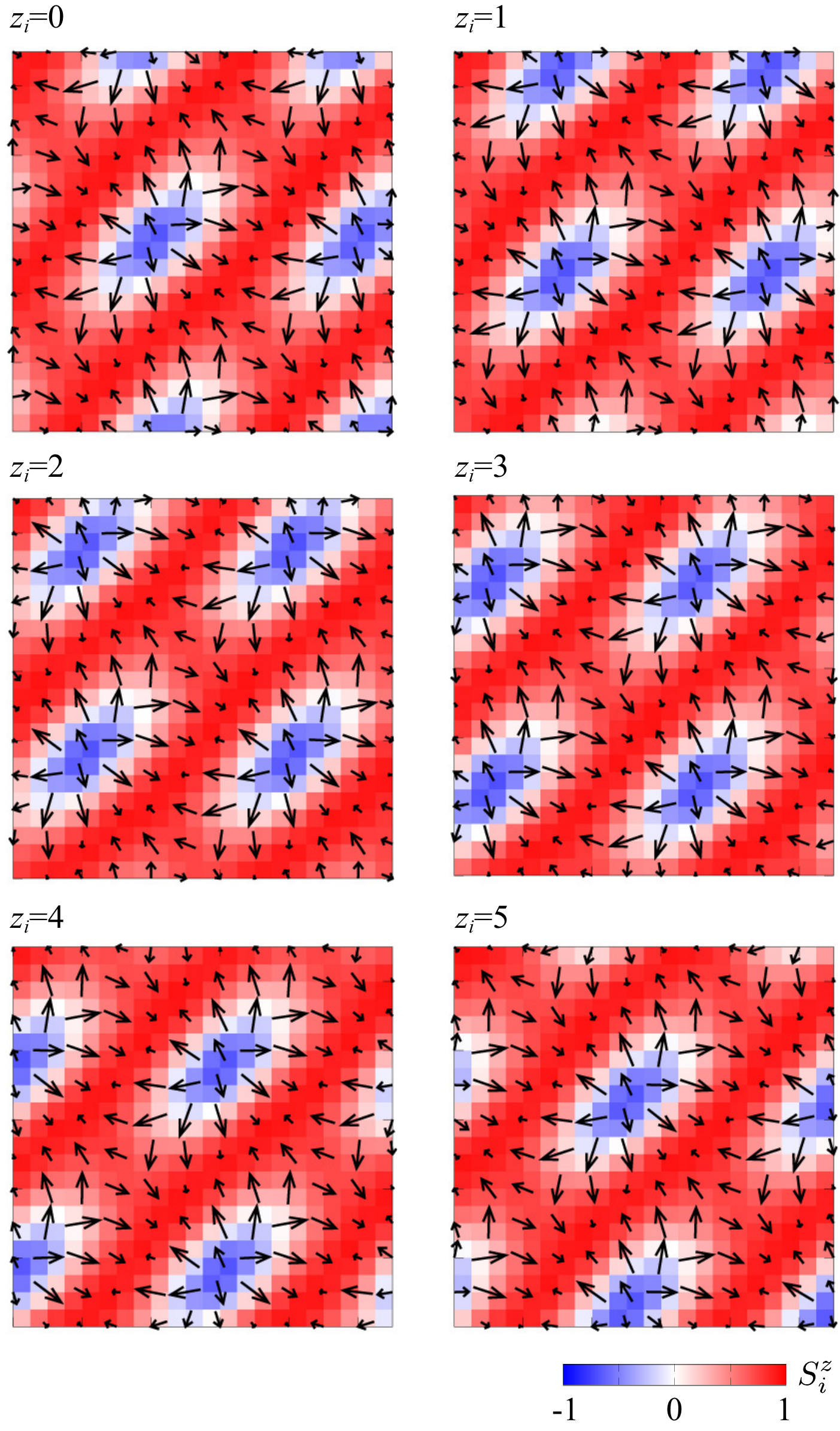} 
\caption{
\label{fig: spin_skx3} 
Real-space spin configuration of the 3$Q'$ SkX at $\kappa=0.9$ and $H=1.3$ for $\bm{H} \parallel [001]$ on the $xy$ plane at $z_i=0$--$5$. 
The arrows represent the in-plane spin components, while the color represents the out-of-plane spin component.  
The spin configuration is drawn by the single snapshot at the lowest temperature.
}
\end{center}
\end{figure}

\begin{figure}[t!]
\begin{center}
\includegraphics[width=1.0\hsize]{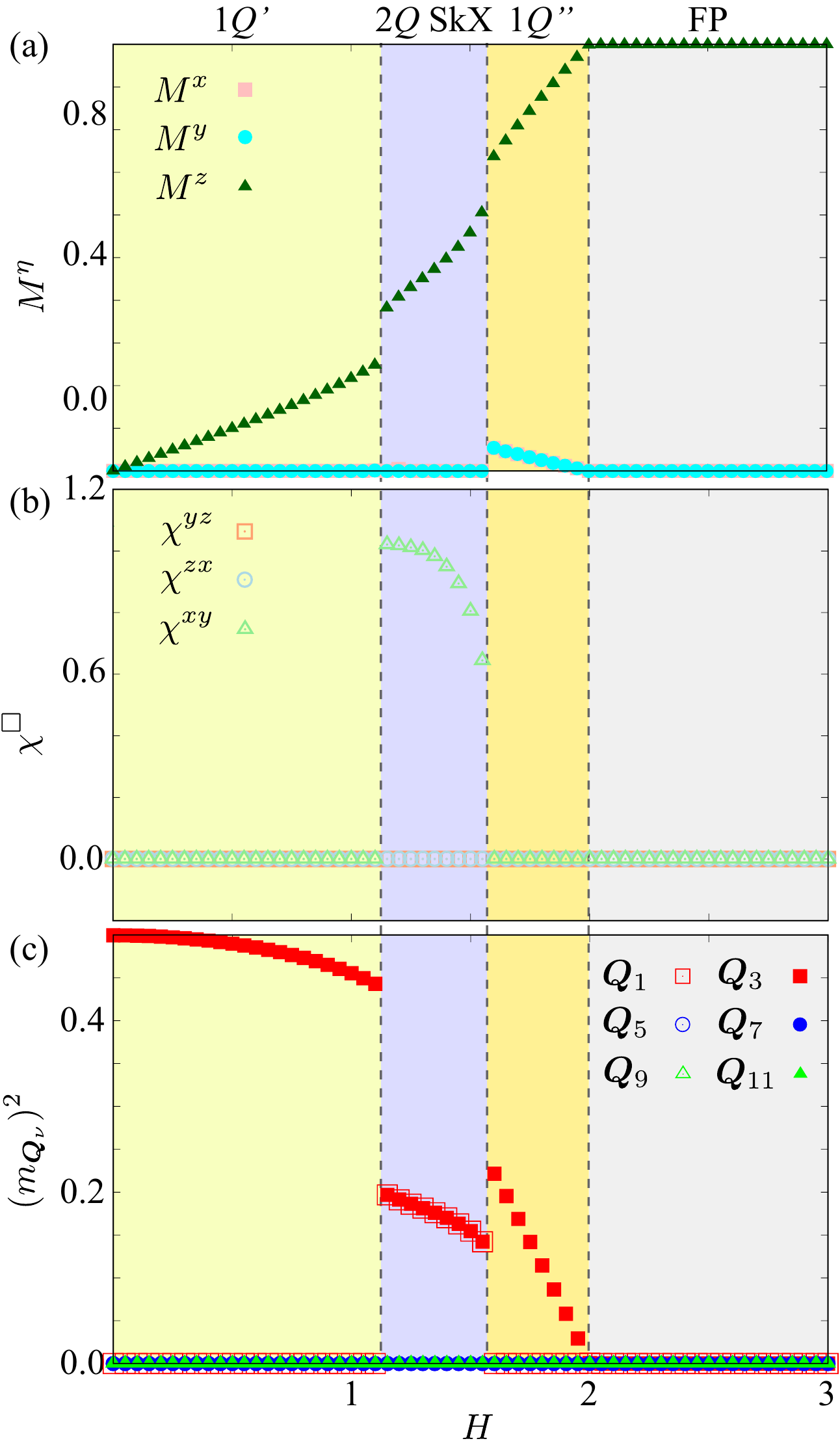} 
\caption{
\label{fig: mag_001_dist_2} 
$H$ dependence of (a) the magnetization $M^\eta$ for $\eta=x,y,z$, (b) the scalar chirality $\chi^{\square}$ for $\square=yz, zx, xy$, and (c) $(m_{\bm{Q}_{\nu}})^2$ for $\nu=1,3,5,7,9,11$ at $\kappa=0.8$ for the model in Eq.~(\ref{eq: Ham_D2d}) under the [001] field. 
The vertical dashed lines show the phase boundaries. 
}
\end{center}
\end{figure}

\begin{figure}[t!]
\begin{center}
\includegraphics[width=0.5\hsize]{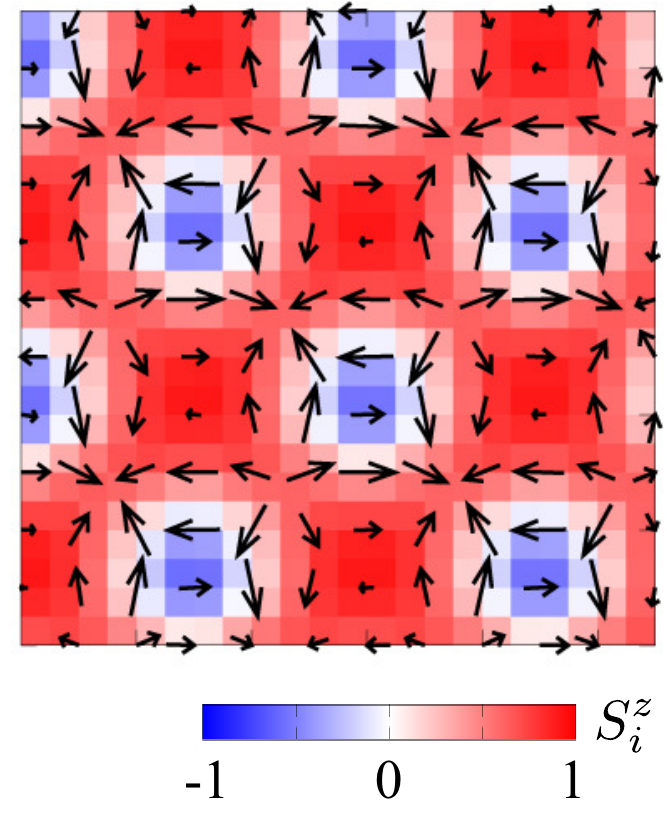} 
\caption{
\label{fig: spin_skx2} 
Real-space spin configuration of the 2$Q$ SkX at $\kappa=0.8$ and $H=1.2$ for $\bm{H} \parallel [001]$ on the $xy$ plane at $z_i=0$.
The arrows represent the in-plane spin components, while the color represents the out-of-plane spin component.  
The spin configuration is drawn by the single snapshot at the lowest temperature.
}
\end{center}
\end{figure}

Figure~\ref{fig: PD2} shows the phase diagram obtained by the simulated annealing while $\kappa$ and $H$ are varied. 
The result for $\kappa=1$ corresponds to that in Fig.~\ref{fig: PD}(d). 
By introducing the effect of uniaxial anisotropy $\kappa<1$, two types of SkXs are realized in the intermediate-field region: 3$Q'$ SkX and 2$Q$ SkX. 
We show the details of the two SkXs below. 

First, we discuss the 3$Q'$ SkX stabilized for $0.85 \lesssim \kappa \lesssim 0.94$, which appears in the region between the 1$Q$ and 1$Q'$ states; the $1Q'$ state is characterized by the single-$Q$ cycloidal spiral state at $\bm{Q}_1$--$\bm{Q}_4$. 
The phase transition between the 3$Q'$ SkX and $1Q$ $(1Q')$ state is discontinuous, as found in the magnetization in Fig.~\ref{fig: mag_001_dist_1}(a). 
The real-space spin configuration of the 3$Q'$ SkX is shown in Fig.~\ref{fig: spin_skx3}. 
In each $z$-coordinate, the SkX with the distorted skyrmion core forms the square lattice, which is similar to the 3$Q$ SkX under the [111] magnetic field in Fig.~\ref{fig: spin}. 
In fact, this state exhibits nonzero spin scalar chirality and triple-$Q$ peaks at $\bm{Q}_\nu$, as shown in Figs.~\ref{fig: mag_001_dist_1}(b) and \ref{fig: mag_001_dist_1}(c), respectively, which is also a similar tendency to the $3Q$ SkX in Fig.~\ref{fig: mag_111}.
On the other hand, there are two main differences between the 3$Q'$ SkX and the 3$Q$ SkX: 
One is the degeneracy of the SkXs.
In the case of the 3$Q$ SkX in Fig.~\ref{fig: mag_111}, the state with $\chi^{xy}, \chi^{yz}, \chi^{zx}<0$ is always selected, since the choice of the triple-$Q$ ordering vectors is unique in order to satisfy $\bm{Q}_\nu \perp \bm{H}$.
Meanwhile, the plane consisting of the triple-$Q$ ordering vectors in the 3$Q'$ SkX is not perpendicular to the field direction in the present [001]-field case. 
For example, the triple-$Q$ ordering vectors $\bm{Q}_3$, $\bm{Q}_5$, and $\bm{Q}_{10}$ in Fig.~\ref{fig: mag_001_dist_1}(c) are perpendicular to the $[\bar{1}\bar{1}1]$ axis rather than the [001] axis while keeping $\bm{Q}_3+\bm{Q}_5+\bm{Q}_{10}=\bm{0}$. 
In other words, the plane consisting of the triple-$Q$ ordering vectors is tilted from the plane perpendicular to the field. 
Such a tilting causes uniform negative magnetizations on the $xy$ plane, i.e., $M^x<0$ and $M^y<0$, as shown in Fig.~\ref{fig: mag_001_dist_1}(a), which results in the positive scalar chirality of $\chi^{yz}$ and $\chi^{zx}$ in Fig.~\ref{fig: mag_001_dist_1}(b). 
Accordingly, $m_{\bm{Q}_3}$ has a different amplitude from $m_{\bm{Q}_5}$ and $m_{\bm{Q}_9}(=m_{\bm{Q}_{10}})$, as shown in Fig.~\ref{fig: mag_001_dist_1}(c). 
Similarly, the other three types of the 3$Q'$ SkX are possible as energetically degenerate states by selecting the different triple-$Q$ superposition so as to be perpendicular to the $[111]$, $[1\bar{1}\bar{1}]$, or $[\bar{1}1\bar{1}]$ axis; the sign of $M_x$ ($\chi_{yz}$) and $M_y$ ($\chi_{zx}$) can be reversed for the different axis while the sign of $M_z$ ($\chi_{xy}$) is fixed by the [001] field. 
Another difference is found in the real-space spin configuration in Fig.~\ref{fig: spin_skx3}; the N\'eel-type helicity around the skyrmion core is realized in the $3Q'$ SkX in contrast to the Bloch-type one in the 3$Q$ SkX in Fig.~\ref{fig: spin}.
This is owing to the different nature of the DM interaction; the triple-$Q$ superposition of the cycloidal spirals is favored in the case of $T_{\rm d}$, while that of the proper-screw spirals is favored in the case of $(O,T)$.

With a further small $\kappa$, the 2$Q$ SkX appears instead of the 3$Q'$ SkX shown in Fig.~\ref{fig: PD2}. 
This state appears as the magnetic field increases in the 1$Q'$ state for $\kappa \lesssim 0.88$, whose transition is identified by the jump of $M^z$ in Fig.~\ref{fig: mag_001_dist_2}(a). 
The real-space spin configuration at $z_i=0$ is shown in Fig.~\ref{fig: spin_skx2}, which remains the same for the other $z$-coordinate; there is no contribution from $\bm{Q}_5$--$\bm{Q}_{12}$. 
In contrast to the other SkXs in Figs.~\ref{fig: spin}, \ref{fig: spin_skx6}, and \ref{fig: spin_skx3}, this state is characterized by the anti-skyrmion structure with positive scalar chirality, i.e., $\chi^{xy}>0$, as shown in Fig.~\ref{fig: mag_001_dist_2}(b). 
This is naturally understood from the DM interaction at the point group $D_{\rm 2d}$~\cite{nayak2017discovery, peng2020controlled}, although the stability region of the SkX is similar to the polar point group $C_{\rm 4v}$ with the different sign of the DM interaction~\cite{Hayami_PhysRevLett.121.137202}.  
Owing to the uniform structure along the $z$ direction, scalar chirality across the different $xy$ planes vanishes $\chi^{yz}=\chi^{zx}=0$ [Fig.~\ref{fig: mag_001_dist_2}(b)]. 
In momentum space, the 2$Q$ SkX is expressed as a superposition of the double-$Q$ spiral states at $\bm{Q}_1$ and $\bm{Q}_3$ with the same intensity, $m_{\bm{Q}_1}=m_{\bm{Q}_3}$, as shown in Fig.~\ref{fig: mag_001_dist_2}(c). 

The 2$Q$ SkX changes into the 1$Q''$ state for small $\kappa$ and the other magnetic (OM) states for large $\kappa$ in Fig.~\ref{fig: PD2}; the 1$Q''$ state is almost characterized by the conical spiral structure with the spiral plane on the $xy$ plane at $\bm{Q}_1$ or $\bm{Q}_3$, although there is a small contribution in the $z$-spin component by the DM interaction, which leads to the uniform in-plane magnetization perpendicular to $\bm{Q}_\nu$. 
The OM state is characterized by a transient multiple-$Q$ state between the 1$Q$ and 1$Q''$ states.

\section{Summary and discussion}
\label{sec: Summary}

To summarize, we have investigated the low-temperature magnetic phase diagrams of the spin model in noncentrosymmetric cubic point groups, $(O,T)$ and $T_{\rm d}$. 
By performing the simulated annealing, we systematically examined the instability toward the multiple-$Q$ states in $(O,T)$ and $T_{\rm d}$. 
Reflecting the difference of the DM interactions between $(O,T)$ and $T_{\rm d}$, the SkX is realized only for the point group $(O,T)$; the SkX corresponds to the triple-$Q$ structure (sextuple-$Q$ structure) for the [111] ([001]) magnetic field. 
The difference of the SkX instability between $(O,T)$ and $T_{\rm d}$ is understood from the different orientations of the DM vector, where the stabilization tendency is consistent with Lifshitz invariants in the free energy~\cite{dzyaloshinskii1964theory,kataoka1981helical, Bogdanov89, Bogdanov94}. 
In other words, our results answer why the SkX has not been observed in materials under the $T_{\rm d}$ symmetry~\cite{Tokura_doi:10.1021/acs.chemrev.0c00297}. 
Meanwhile, we have shown that the triple-$Q$ and double-$Q$ SkXs appear under $T_{\rm d}$ by introducing the uniaxial strain lowering the symmetry to $D_{\rm 2d}$. 
The above results indicate the emergence of various types of SkXs for the different lattice structures and the magnetic-field direction. 
Since several lattice structures, such as a zinc-blende structure, cubic half-Heusler structure, A-site cation-ordered spinel structure in LiFeCr$_4$O$_8$~\cite{Saha_PhysRevB.96.214439}, and other cubic structures like Ho(In,Cd)Cu$_4$~\cite{Stockert_PhysRevResearch.2.013183}, has the $T_{\rm d}$ symmetry, the SkX might be expected by applying external pressure.

Finally, let us discuss the possibilities of the SkX under the point group $T_{\rm d}$ without uniaxial strain. 
One is to introduce the anisotropic exchange interaction that is neglected in the model in Eq.~(\ref{eq: Ham1}). 
In the present ordering vectors for $\bm{Q}_\nu \parallel  \langle 110 \rangle$, there are two types of anisotropic exchange interactions~\cite{yambe2023anisotropic}: One is $ S^x_{\bm{Q}_1}   S^x_{-\bm{Q}_1}=S^y_{\bm{Q}_1}  S^y_{-\bm{Q}_1} \neq S^z_{\bm{Q}_1}   S^z_{-\bm{Q}_1}$ and the other is $S^x_{\bm{Q}_1}   S^y_{-\bm{Q}_1}+S^y_{\bm{Q}_1}   S^x_{-\bm{Q}_1}$ for $\bm{Q}_1$. 
Since these interactions tend to deform the circular shape of the spiral plane leading to the energy cost by higher-harmonic contributions, they can become a source of inducing multiple-$Q$ states~\cite{amoroso2020spontaneous, Hayami_PhysRevB.103.054422, yambe2021skyrmion}. 
Another is the multiple-spin interaction like the form of $K(\bm{S}_{\bm{Q}_\nu} \cdot \bm{S}_{-\bm{Q}_\nu})^2$ with the positive coupling constant $K>0$.  
This interaction also becomes the origin of multiple-$Q$ states irrespective of the presence/absence of the DM interaction~\cite{Hayami_PhysRevB.95.224424, Okumura_PhysRevB.101.144416, hayami2021field, Okumura_doi:10.7566/JPSJ.91.093702}. 
Indeed, we obtained a similar spin configuration to the 3$Q'$ SkX under the [001] field at $K=0.4$ and $D=0.3$ (not shown).

\begin{acknowledgments}
This research was supported by JSPS KAKENHI Grants Numbers JP21H01037, JP22H04468, JP22H00101, JP22H01183, JP23H04869, JP23K03288, and by JST PRESTO (JPMJPR20L8). 
R.Y. was supported by Forefront Physics and Mathematics Program to Drive Transformation (FoPM).
Parts of the numerical calculations were performed in the supercomputing systems in ISSP, the University of Tokyo.
\end{acknowledgments}

\bibliography{ref}

\begin{thebibliography}{49}
\bibitem [{com()}]{comment_polarDM}%
  \BibitemOpen
  \href@noop {} {}\bibinfo {note} {It is noted that this type of DM interaction
  has the same symmetry as the octupole $xyz$ in real space instead of the
  dipole $\bm{r}$ owing to the presence of threefold rotational symmetry around
  the $\langle 111\rangle$ axis.}\BibitemShut {Stop}%
\end{thebibliography}
\end{document}